\documentclass[12pt,preprint]{aastex}
\usepackage{epsfig,natbib}
\newcommand{\tstd}{\tau_{\rm std}}
\newcommand{\HH}{H$_{2}$}
\newcommand{\bi}{$b_{i}$}

\shorttitle{NLTE Irradiated Model}
\shortauthors{Barman et al.}

\begin{document}
\bibliographystyle{apj}

\title{Non-LTE Effects of Na I in the Atmosphere of HD209458b}

\author{Travis S. Barman, Peter H. Hauschildt, Andreas Schweitzer, Phillip C. Stancil}
\affil{Dept.\ of Physics and Astronomy \& Center for Simulational Physics, 
       University of Georgia, Athens, GA 30602-2451, USA\\}
\and
\author{E. Baron}
\affil{Dept.\ of Physics and Astronomy, University of Oklahoma, 440 W. 
       Brooks, Rm 131, Norman, OK 73019-0260, USA\\ } 
\and
\author{France Allard}
\affil{C.R.A.L (UML 5574) Ecole Normale Superieure, 69364 Lyon Cedex 7, France\\}

\begin{abstract}
The recent announcement that sodium absorption has been observed in the
atmosphere of HD209458b, the only EGP observed to transit its parent star, is
the first direct detection of an EGP atmosphere.  We explore the possibility
that neutral sodium is {\em not} in local thermodynamic equilibrium (LTE) in
the outer atmosphere of irradiated EGPs and that the sodium concentration may
be underestimated by models that make the LTE assumption.  Our results indicate
that it may not be necessary to invoke excessive photoionization, low
metallicity, or even high altitude clouds to explain the observations.
\end{abstract}


\section{Introduction}

HD209458b is the first extra-solar giant planet (EGP) observed to transit
its parent star \cite[]{Charbon00,Henry00} and, consequently, its mass
(0.69 M$_{\rm Jupiter}$) and radius (1.35 R$_{\rm Jupiter}$) are known to a
high degree of accuracy.  These results leave little doubt that HD209458b
is a gas giant.  and its special orbital inclination has drawn a great deal
of attention from observers and theorists over the past year.  Recently, an
increase in the sodium absorption (relative to the continuum) at 5893\AA\
was observed with the Hubble Space Telescope (HST) during several transits
of HD2094598 \cite[]{Charbon01}.  The additional sodium absorption is
believed to be due to Na D absorption in the EGP's atmosphere as stellar
light passes through the planetary limb.  This observation marks a major
turning point in the study of EGPs, for now we have direct evidence of an
atmosphere around HD209458b and a measurement of one chemical species (Na).
According to \cite{Charbon01}, the HST observations suggest that either the
EGP atmosphere has a low concentration of neutral atomic Na (due to
photoionization, molecular formation, or an overall low metallicity) or
that high altitude clouds exist and reduce the amount of stellar flux
transmitted though the EGP limb. 

In this letter, we offer an alternate explanation for the observed Na
absorption and explore the possibility that the Na D feature is altered by
nonlocal thermodynamic equilibrium (NLTE) effects brought on by the impinging
stellar radiation field and insufficient collisional thermalization.  If true,
NLTE effects would offer a natural explanation for the apparently low sodium
absorption observed in the HD209458b without the need for excessive ionization,
a reduced metallicity, or extremely high altitude clouds.  Below we will
present theoretical predictions for NLTE Na D doublet line profiles for the
transmitted spectrum of an irradiated EGP atmosphere and provide stringent
limits on the non-LTE effects.

\section{Model Construction}

The construction of the model atmospheres presented below follows the procedure
outlined in \cite{Barman01} (here after BHA).  In order to make a direct
comparison with HD209458b, we have adopted the best-fit values for the radius,
mass, and orbital separation of HD209458b and its parent star published by
\cite{Brown012} and \cite{Mazeh} (see table \ref{tab1}).  We have also assumed
that the total flux received by the planet has been uniformly distributed over
the planet's dayside.  Therefore, the incident flux has been weighted by 0.5,
whereas in the BHA models, no redistribution was assumed. The intrinsic
effective temperature of the model (i.e. the effective temperature of the
planet in the absence of irradiation) is 500K. However, this intrinsic
temperature has little effect on the structure of the outer atmosphere which is
completely determined by the incident flux.  With this redistribution, our
models represent an upper limit to the equilibrium effective temperature which
is about 1800K.  We further assume that the planet has a solar composition with
an opacity setup identical to the ``AMES-cond'' models of BHA and
\cite{Allard01}.  In this situation, dust grains form in the atmosphere at
locations determined by the chemical equilibrium equations but their opacity
contribution is ignored, mimicking a complete removal of the grains by
efficient gravitational settling.  Therefore, the models in this study
represent cloud free atmospheres.  
  
We model the flux transmitted through the limb of the planet's
atmosphere, by solving the spherically symmetric radiative transfer equation
(SSRTE) suitably adjusted to account for the incident radiation.  The incident
flux is taken from a separate calculation which reproduces the observed
spectrum of HD209458.  The SSRTE has a significant advantage over the
plane-parallel solution because a geometry more appropriate (and more accurate)
for the upper atmosphere is already incorporated.  When solving the SSRTE, the
atmosphere is modeled as a discrete number of concentric shells surrounding the
interior (or ``core'').  The solution along characteristic rays that pass
through the outer most shells is all that is needed to calculate the
transmitted intensities and automatically accounts for the curvature of the
limb.  The total transmitted flux is obtained by integrating over the planet
limb:
\begin{equation}
F_{trans,\lambda} = \int_{R_{\rm min}}^{R_{\rm max}} I_{\lambda}(r)2\pi r dr
\end{equation}
where $r$ is the perpendicular distance from the planet center to a 
tangential characteristic ray.  $R_{\rm max}$ is the planet radius at $\tstd =0$
and $\tstd$ is the optical depth at $1.2\micron$.
$R_{min}$ is chosen such that for $r \ge R_{\rm min}$ and $\lambda = 5880\rm\AA$,
$I(r) \ge 0.1 I^{\star}(r)$
where $I^{\star}$ is the incident stellar intensity.  This lower limit
 ensures that the transmitted intensities are well sampled
and light from the core region is excluded.  Note that both $R_{\rm max}$ and
$R_{\rm min}$ are determined by the simulation and depend on the resulting
structure of the model atmosphere.  The only prescribed radius is $R_{\rm p} =
r(\tstd = 1)$, which is equal to the planet radius given in table \ref{tab1}.
With these definitions, the thickness of the
limb ($H = R_{\rm max} - R_{\rm min}$) is roughly $0.07 R_{\rm p}$.

\subsection{Na in non-LTE}

When a gas is assumed to be in LTE, the level populations for each species
depend entirely upon the gas temperature and electron pressure and are
given by the Saha-Boltzmann distribution.  In general, the LTE assumption
is a matter of computational convenience and is not expected to be valid in
most cases, especially in the optically thin regions of an atmosphere.
However, it is often assumed that LTE is achieved in very late type
stars (even Brown dwarfs) despite the fact that this assumption has not
been thoroughly tested.  Departures from LTE have previously been
investigated for Ti I \cite[]{yeti97} and CO \cite[]{Schweitzer00} for cool
M dwarfs and for CH$_4$ in the Jovian planets \cite[]{Appleby90}.  These
earlier works found that, for these particular species, NLTE effects were
small.  

Due to the close proximity of HD209458b to its parent star, the planet's
atmosphere is subjected to intense stellar radiation.  This inherently {\em
nonlocal} source of radiation dramatically alters the conditions of the outer
atmosphere compared to an isolated EGP.  In addition, the cool atmospheres of
EGPs are dominated by strong opacity sources like H$_2$O and CH$_4$ and,
therefore, have intensities that differ greatly from those of a blackbody.
When combined with the relatively low pressures of the outer atmosphere, these
conditions very likely will lead to departures from LTE which may play an
important role in determining the atmospheric structure and resulting spectrum.
If LTE is to be achieved in EGP atmospheres, then the collisional rates must be
large enough to compensate for the deviations of the radiative rates from their
LTE values.  Unfortunately, a limitation of all NLTE models is a lack of well
determined collisional cross-sections for interactions with important species
like \HH, He, and H.  However, it is unlikely, given the low thermal
velocities, that these particles could restore LTE completely.  In fact, the
effects of collisions with hydrogen on the Na D profile in M dwarfs with
chromospheres are fairly small and electron collisions dominate despite the
fact that $N_{H}/N_{e} \sim 10^6$ \cite[]{Andretta97}. 

The Na I model atom used in this work includes 53 levels, and 142 primary
transitions (all bound-bound transitions with $\log(gf) > -3.0$) were included
in the solution of the statistical equilibrium equations.  For details of the
method used to solve the rate equations, see \cite{yeti99}.  As in
\cite{yeti97} for Ti I, electron-impact bound-free collisional rates are
approximated by the formula of \cite{drawin61} and bound-bound collisional
rates are based on the semi-empirical formula of \cite{allen_aq} with permitted
transitions determined from Van Regemorter's (1962)\nocite{vr62} formula.  For
the ground-state photoionization cross sections, we have used the Opacity
Project data of \cite{Bautista98}.  We have also assumed complete
redistribution.

We have constructed four model atmospheres and resulting transmitted spectra
for HD209458b. The first irradiated model (A) only includes collisions with
electrons, where the source of free electrons is primarily the ionization of
potassium.  Even in an irradiated EGP atmosphere, the number density of
electrons is very small ($N_{e}/N_{\rm H2} \sim 10^{-8}$) and the temperatures
are too low for electronic collisions to be important.  Model A, therefore, can
be considered a lower limit for the collisional rates.  The second irradiated
model (B) has the same parameters as model A, except that collisions with \HH\
are also included.  In order to place a secure upper limit on the effects of
collisions with \HH, we have treated \HH\ as if it had the same rate coefficients
as an electron.  Assuming identical cross-sections implies that the rate
coefficients scale with $(\mu)^{-\frac{1}{2}}$, where $\mu$ is the reduced mass
of the collision partners.  Therefore, we are overestimating the \HH\
collisional rates by more than an order of magnitude.  The remaining two models
(C and D) represent {\em non}-irradiated atmospheres with the same effective
temperature as the irradiated models ($\sim$ 1800K) and include the same lower
(model C) and upper (model D) limits for the collision rates.  Each of these
model atmospheres was produced from a self-consistent solution of the SSRTE,
chemical equilibrium equations, and the NLTE rate equations. We did not
prescribe the temperature structure or mixing ratios for any of the species. 

\section{Results}
\begin{figure*}
\plotone{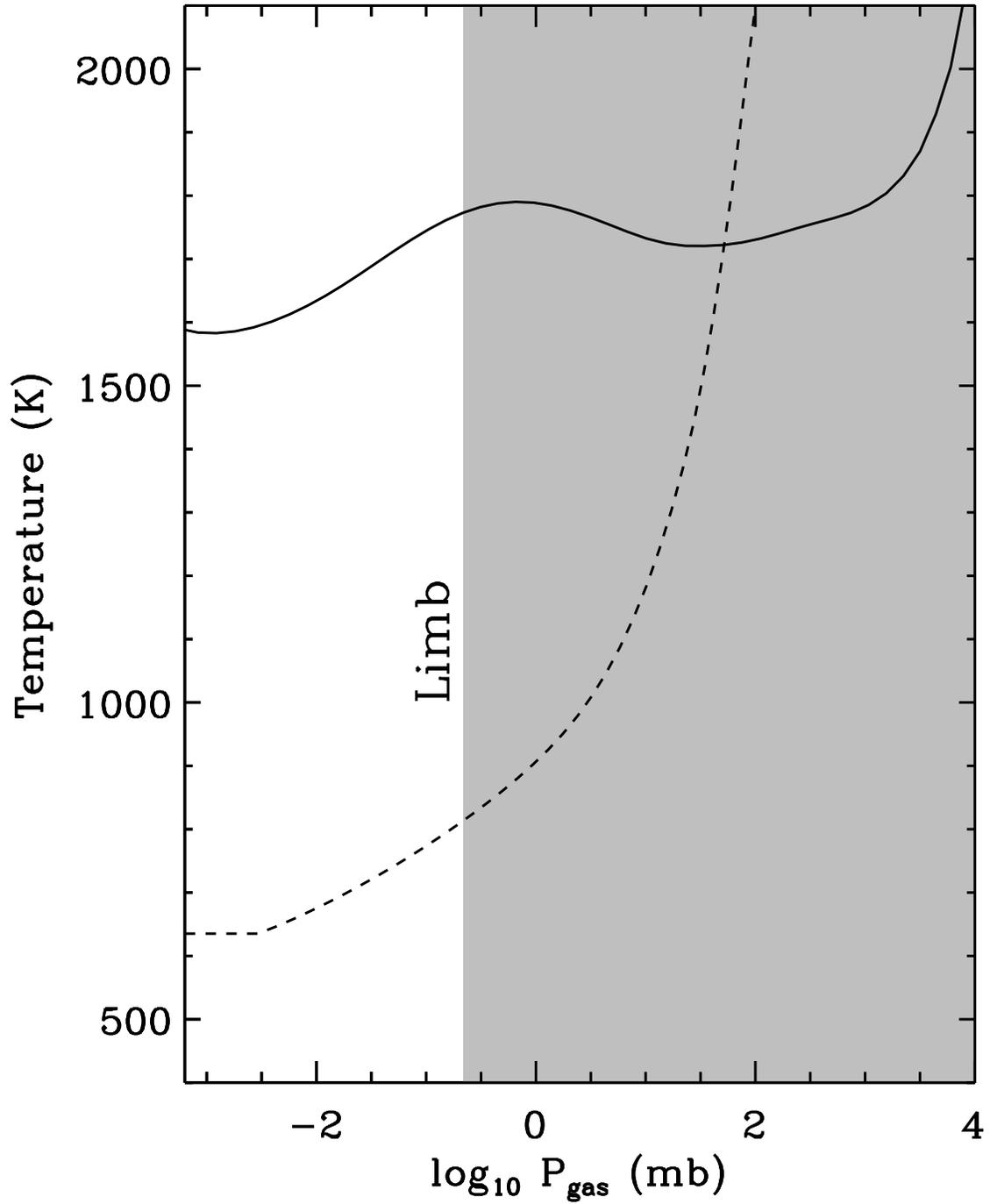}
\caption{
The temperature-pressure profiles for the irradiated (solid line) and 
non-irradiated (dashed line) models.  The unshaded area indicates the
limb region for the irradiated models.
\label{fig1}}
\end{figure*}

The temperature-pressure profiles for models A and C are shown in figure
\ref{fig1}.  Note that, the profile for model B is identical to that of model A
and model D has the same profile as model C.  The gas pressures in the limb
are very low, ranging from 0.004mb to 0.2mb.  As is to be expected, the
non-irradiated models look nothing like the irradiated models which have
temperatures nearly 1000K higher in the outer atmosphere
\cite[]{Barman01,Gouken2000,Seager1998}.  There were no significant changes to
the temperature-pressure profiles in the four NLTE models compared to
the LTE structures.  

\begin{figure*}
\plotone{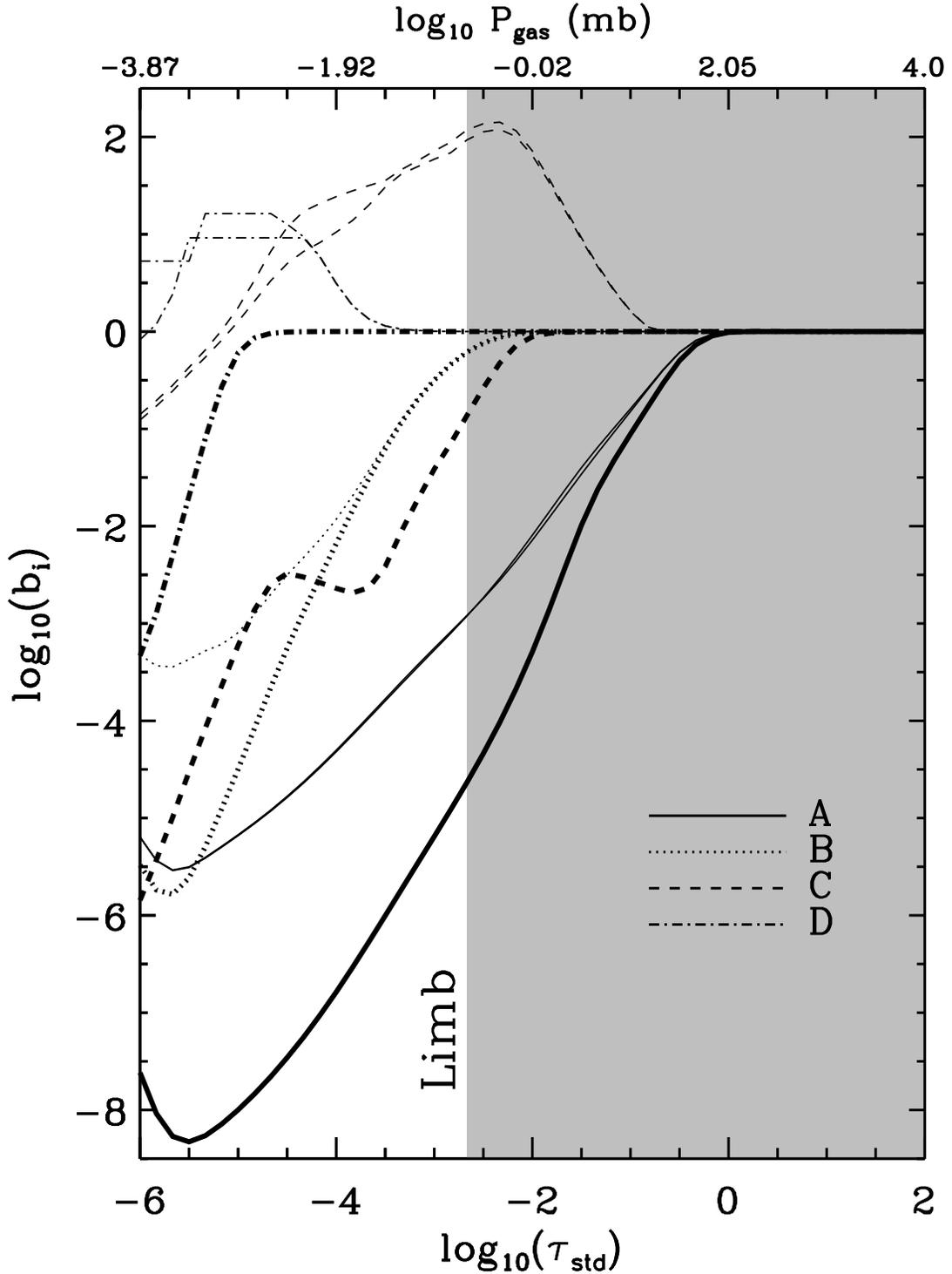}
\caption{
The departure coefficients for neutral sodium in the
atmosphere of HD209458b as a function of the radial optical depth at 1.2\micron
($\tstd$) and pressure (top axis).  Model A demonstrates departures for an
irradiated atmosphere including only collisions with free electrons.  Model B
is the same as A, but includes collisions with \HH\ assuming that \HH\ has the
same rate coefficients as an electron.  Model C shows the departures in a
non-irradiated atmosphere (where $T_{eq} = T_{\rm eff}$) with collisions
treated as in Model A. Model D is the same as C but with collisions treated as
in Model B. Thick lines refer to the 3s level (ground state) and thin lines
indicate the 3p levels (for $J = 1/2$ and $J = 3/2$).  The unshaded region
is the portion of the atmosphere where stellar flux is transmitted 
(Limb) in the irradiated case.  
\label{fig2}}
\end{figure*}

Departures of the level populations from LTE, for a particular species, are
usually described by the departure coefficients, $b_i=n_{i}^{\star}/n_{i}$,
where $n_{i}^{\star}$ is the NLTE population density for level $i$ and $n_{i}$
is the LTE value \cite[]{mihalas1}.  In figure \ref{fig2}, we show the \bi's
for the ground state (3s) and the first excited states (3p$_{\frac{1}{2}}$,
3p$_{\frac{3}{2}}$) of the neutral sodium atom for several different physical
conditions within an EGP atmosphere.  The largest departures are seen in model
A where the 3s (ground state) and 3p levels are both underpopulated by many
orders of magnitude, especially in the limb region.  The main reason for such
large departures is that we have only included collisions with electrons and,
in such a cool atmosphere, the number density of electrons is $\sim 8$ orders
of magnitude below that of the dominant species, \HH.  Consequently, there are
essentially no collisions in model A to thermalize the level populations in the
Na atom thus allowing the radiative rates to dominate and drive the system out
of LTE.  Also, since the upper atmosphere is dominated by a large external
radiation source (which happens to peak near the Na D doublet, $\sim 5000$\AA),
the mean intensity of the line is much larger than the thermal source function
which implies a strong decoupling of the radiation field from the local
conditions.  Furthermore, the ratio of the line source function to a blackbody
is roughly given by the ratio of the departure coefficients of the upper (3p)
and lower (3s) levels for the transition \cite[]{Bruls92}.  From figure 2, we
see that the line source function is far from a blackbody for the majority of
the upper atmosphere (since $b_{3p}/b_{3s} \gg 1$).

An obvious question to ask is whether the system will return to LTE if
collisions with the dominant species (\HH) are included.  Model B, which
has the same parameters and temperature profile as model A, includes
collisions with both electrons {\em and} \HH\ (but assuming that \HH\ has
the same rate coefficients as an electron).  In effect, we have increased the
importance of electronic collisions by more than 8 orders of magnitude.  As
is expected, the departure coefficients are closer to one, but the level
populations in the limb are still far from the LTE values. However, since
$b_{3p}/b_{3s} \sim 1$ for most of the atmosphere, NLTE effects will
only be important in model B for $\tstd < 10^{-4}$.  The real situation is
likely to be somewhere between models A and B.

We also show departure coefficients for non-irradiated models (C and D) with
$T_{\rm eff}$ equal to the equilibrium effective temperature of the irradiated
atmospheres (1800K) for both collisional rate limits.  Despite the absence of
any extrinsic radiation and much lower temperatures in the upper layers, the
non-irradiated atmospheres still have departures from LTE, though generally
less significant than those found in the irradiated atmospheres.  Model C (the
near collision free limit), has departures from LTE similar to those of the
collision dominated irradiated model (B) for the ground state but with an over
populated 3p level.  With increased collisional rates, Na I has nearly returned
to LTE for most of the atmosphere in model D with departures still present in
the top most layers whereas in the irradiated case both levels were greatly
underpopulated even in the collision dominated model (B).

\begin{figure*}
\plotone{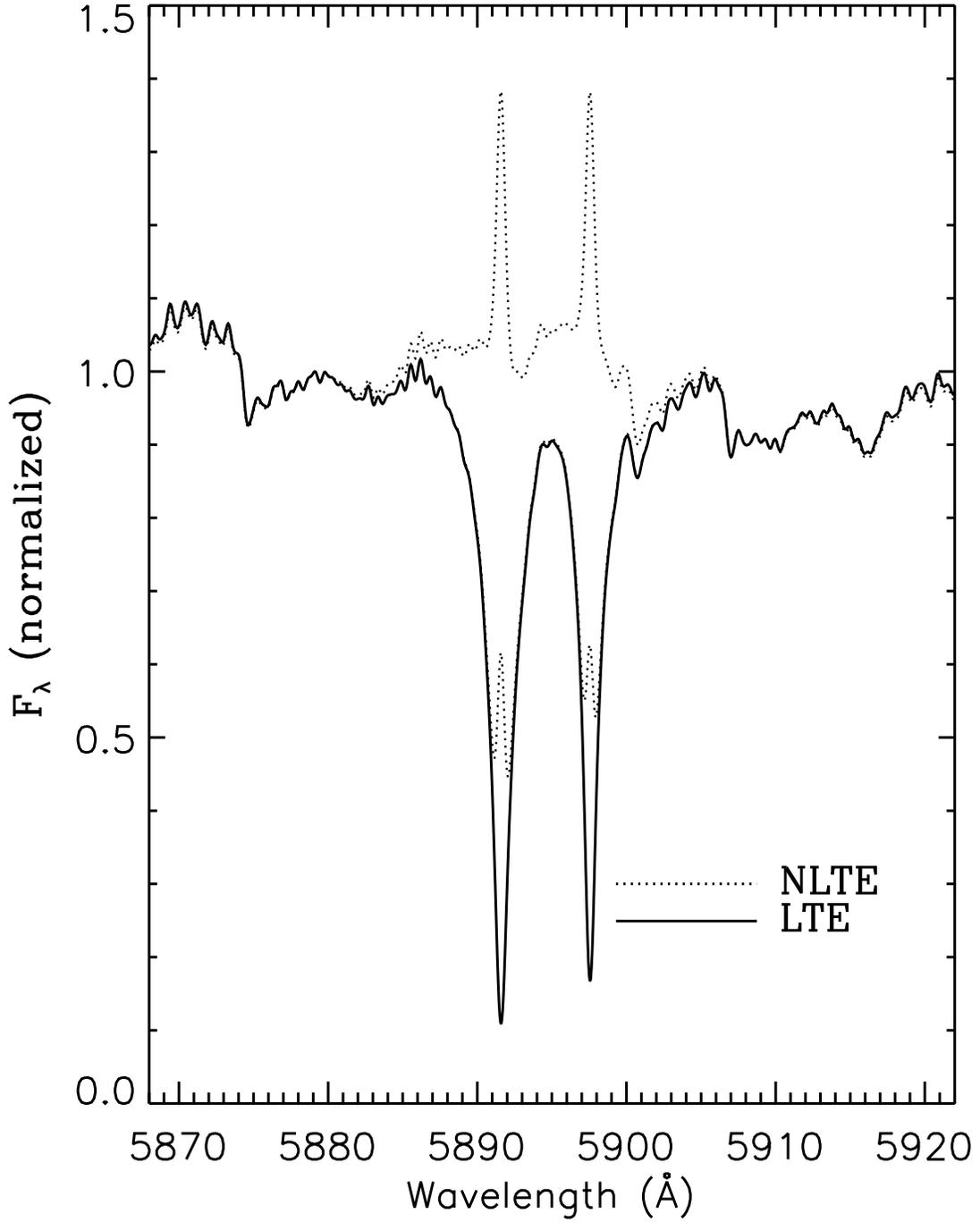}
\caption{
The Na D doublet for Model A (top dotted line), Model B (lowest dotted
line), and when assuming LTE (solid line).  The flux has been normalized to
one at 5880\AA.
\label{fig3}}
\end{figure*}

The effects on the Na D line profiles are quite dramatic  for model A (see
figure \ref{fig3}).  In this case, the lack of thermalization reduces the line
transfer to nearly a pure scattering case.  As a result, the doublet appears
completely in emission.  However, in model B, as the collisional rates are
increased, the line wings return to their LTE shape while the line cores are
reversed.  The data analysis of \cite{Charbon01} does not directly reveal the
Na line profile produced by the planet atmosphere.  Instead, their work shows
that a deeper transit is observed in the Na band implying additional Na
absorption by the planet limb.  Given the current sensitivity of the
observations, a core reversal feature like the one shown in our model B would
be buried in the noise and manifest itself as simply a smaller equivalent
width.  The Na absorption in model B would result in a transit deeper than in
the continuum bands but not as deep as implied by the LTE model.  The fact that
Na D absorption (and not emission) has been observed in the transmitted
spectrum of HD209458b rules out our model A indicating that some thermalization
does occur in the limb region.  However, it is unlikely that the collisional
rates are as large as those in our model B suggesting that the equivalent width
of the Na D doublet will be substantially reduced by NLTE effects. 

The reduced equivalent width predicted by our model is {\em not} due to
photoionization of Na.  In the majority of the limb, only 3\% is ionized
and the neutral Na concentration is nearly constant with $N_{\rm Na}/N_{\rm
H2} \sim 10^{-5.5}$.  Na is only significantly ionized ($\gg 5\%$) at the
very top of the atmosphere where $\rm P_{\rm gas} < 1 \mu bar$.  The
shallow ionization depth of Na is due to the strong UV opacity provided by
metals (e.g. atomic Mg, Al, Ca, Fe, and Ni) which effectively shield Na
from the incident ionizing photons.  The ionization predicted by our models
(even 3\%) is far greater than what is obtained from an LTE calculation but
does not significantly affect the line profile.  However, if the planet's
atmosphere is substantially cooler than in our model, then additional
condensation and settling could further deplete the atmosphere of metals
and allow greater ionization of Na to occur.  We also find that only a very
small amount of Na is in molecular form and that neutral Na is nearly 3
orders of magnitude more abundant than NaCl (the most abundant Na bearing
molecule).  Furthermore, condensation of Na via NaCl grains is unlikely.
The reduced equivalent width and central core emission is purely a
radiative transfer effect.  

\section{Conclusions}
The recent HST observations provided the first direct measurement of the
conditions inside the atmosphere of HD209458b.  However, even under the best
circumstances, determining the concentration of any species based solely upon
one absorption feature is problematic, especially  if this feature forms in the
upper regions of an atmosphere where pressures are low and NLTE effects are
greatest.  Our models clearly show that Na is far from being in LTE in the
upper atmosphere of HD209458b and the observed Na absorption can be explained
with a solar metallicity atmosphere which is cloud free or has only very low
lying clouds.  It is likely that other important species (e.g. CO and CH$_4$)
are in NLTE, and we plan to test the LTE assumption for a wide variety of
atomic and molecular species in a future work. Hopefully the
Na D doublets and other alkali metal lines will be useful diagnostics in the
study of EGP atmospheres.  However, only with detailed NLTE calculations
including well determined collisional rates will we have a chance at
constraining the physical conditions in the atmosphere of HD209458b.  

\acknowledgments
We would like to thank Charbonneau et al. for making a copy of their work
available before publication.  This research was supported by NASA ATP and LTSA
grants to the University of Georgia and to Witchita State and in part by the
P\^ole Scientifique de Mod\'elisation Num\'erique at ENS-Lyon.


\clearpage

\begin{deluxetable}{rccrrrrr} 
\tablecolumns{8} 
\tablewidth{0pc} 
\tablecaption{Model Parameters} 
\tablehead{ 
\colhead{}& \colhead{HD209458b}&\colhead{HD209458}}
\startdata 
radius              & 1.35 R$_{\rm Jupiter}$ & 1.146 R$_\odot$ \\
mass                & 0.63 M$_{\rm Jupiter}$ & 1.05 M$_\odot$  \\
T$_{\rm eff}$           & 1800 K     & 6000 K          \\
$[$Fe/H$]$            & 0.00       & 0.00        \\    
semi-major axis &\multicolumn{2}{c}{0.045 AU} \\
\enddata 
\label{tab1}
\end{deluxetable}

\begin{thebibliography}{22}
\expandafter\ifx\csname natexlab\endcsname\relax\def\natexlab#1{#1}\fi

\bibitem[{{Allard} {et~al.}(2001){Allard}, {Hauschildt}, {Alexander},
  {Tamanai}, \& {Schweitzer}}]{Allard01}
{Allard}, F., {Hauschildt}, P.~H., {Alexander}, D.~R., {Tamanai}, A., \&
  {Schweitzer}, A. 2001, \apj, 556, 357

\bibitem[{Allen(1973)}]{allen_aq}
Allen, C.~W. 1973, Astrophysical Quantities, 3rd edn. (London: Athlone Press)

\bibitem[{{Andretta} {et~al.}(1997){Andretta}, {Doyle}, \&
  {Byrne}}]{Andretta97}
{Andretta}, V., {Doyle}, J.~G., \& {Byrne}, P.~B. 1997, \aap, 322, 266

\bibitem[{{Appleby}(1990)}]{Appleby90}
{Appleby}, J.~F. 1990, Icarus, 85, 355

\bibitem[{{Barman} {et~al.}(2001){Barman}, {Hauschildt}, \&
  {Allard}}]{Barman01}
{Barman}, T.~S., {Hauschildt}, P.~H., \& {Allard}, F. 2001, \apj, 556, 885

\bibitem[{{Bautista} {et~al.}(1998){Bautista}, {Romano}, \&
  {Pradhan}}]{Bautista98}
{Bautista}, M.~A., {Romano}, P., \& {Pradhan}, A.~K. 1998, \apjs, 118, 259

\bibitem[{{Brown} {et~al.}(2001){Brown}, {Charbonneau}, {Gilliland}, {Noyes},
  \& {Burrows}}]{Brown012}
{Brown}, T.~M., {Charbonneau}, D., {Gilliland}, R.~L., {Noyes}, R.~W., \&
  {Burrows}, A. 2001, \apj, 552, 699

\bibitem[{{Bruls} {et~al.}(1992){Bruls}, {Rutten}, \& {Shchukina}}]{Bruls92}
{Bruls}, J.~H.~M.~J., {Rutten}, R.~J., \& {Shchukina}, N.~G. 1992, \aap, 265,
  237

\bibitem[{{Charbonneau} {et~al.}(2000){Charbonneau}, {Brown}, {Latham}, \&
  {Mayor}}]{Charbon00}
{Charbonneau}, D., {Brown}, T.~M., {Latham}, D.~W., \& {Mayor}, M. 2000, \apjl,
  529, L45

\bibitem[{{Charbonneau} {et~al.}(2001){Charbonneau}, {Brown}, {Noyes}, \&
  {Gilliland}}]{Charbon01}
{Charbonneau}, D., {Brown}, T.~M., {Noyes}, R.~W., \& {Gilliland}, R.~L. 2001,
  \apj, in press

\bibitem[{Drawin(1961)}]{drawin61}
Drawin, H.~W. 1961, Zs. f. Phys., 164, 513

\bibitem[{{Goukenleuque} {et~al.}(2000){Goukenleuque}, {B{\'e}zard}, {Joguet},
  {Lellouch}, \& {Freedman}}]{Gouken2000}
{Goukenleuque}, C.~., {B{\'e}zard}, B., {Joguet}, B., {Lellouch}, E., \&
  {Freedman}, R. 2000, Icarus, 143, 308

\bibitem[{{Hauschildt} {et~al.}(1997){Hauschildt}, {Allard}, {Alexander}, \&
  {Baron}}]{yeti97}
{Hauschildt}, P.~H., {Allard}, F., {Alexander}, D.~R., \& {Baron}, E. 1997,
  \apj, 488, 428

\bibitem[{{Hauschildt} \& {Baron}(1999)}]{yeti99}
{Hauschildt}, P.~H. \& {Baron}, E. 1999, JCAM, 109, 41

\bibitem[{{Henry} {et~al.}(2000){Henry}, {Marcy}, {Butler}, \&
  {Vogt}}]{Henry00}
{Henry}, G.~W., {Marcy}, G.~W., {Butler}, R.~P., \& {Vogt}, S.~S. 2000, \apjl,
  529, L41

\bibitem[{{Mazeh} {et~al.}(2000){Mazeh}, {Naef}, {Torres}, {Latham}, {Mayor},
  {Beuzit}, {Brown}, {Buchhave}, {Burnet}, {Carney}, {Charbonneau}, {Drukier},
  {Laird}, {Pepe}, {Perrier}, {Queloz}, {Santos}, {Sivan}, {Udry}, \&
  {Zucker}}]{Mazeh}
{Mazeh}, T., {Naef}, D., {Torres}, G., {Latham}, D.~W., {Mayor}, M., {Beuzit},
  J., {Brown}, T.~M., {Buchhave}, L., {Burnet}, M., {Carney}, B.~W.,
  {Charbonneau}, D., {Drukier}, G.~A., {Laird}, J.~B., {Pepe}, F., {Perrier},
  C., {Queloz}, D., {Santos}, N.~C., {Sivan}, J., {Udry}, S.~., \& {Zucker}, S.
  2000, \apjl, 532, L55

\bibitem[{Mihalas(1970)}]{mihalas1}
Mihalas, D. 1970, Stellar Atmospheres, 1st edn. (New York: W.H. Freeman)

\bibitem[{{Schweitzer} {et~al.}(2000){Schweitzer}, {Hauschildt}, \&
  {Baron}}]{Schweitzer00}
{Schweitzer}, A., {Hauschildt}, P.~H., \& {Baron}, E. 2000, \apj, 541, 1004

\bibitem[{{Seager} \& {Sasselov}(1998)}]{Seager1998}
{Seager}, S. \& {Sasselov}, D.~D. 1998, \apjl, 502, L157

\bibitem[{{Van Regemorter}(1962)}]{vr62}
{Van Regemorter}, H. 1962, \apj, 136, 906

\end{thebibliography}
\end{document}